\documentclass[twocolumn,prl,showpacs,amsmath,amssymb,superscriptaddress,nofootinbib,floatfix]{revtex4}
\usepackage{tabularx}
\usepackage{float}
\usepackage{bm}
\usepackage{graphicx}
\usepackage{amsmath}
\usepackage{mathtools}
\usepackage{color}
\usepackage{mathrsfs}
\usepackage{hyperref}

\usepackage{subfiles}

\begin{document}

\title{Broadband enhancement of second harmonic generation at the domain walls of magnetic topological insulators}

\author{G. Rakhmanova}
\author{I.V. Iorsh}
\affiliation{
ITMO University, Kronverkskiy prospekt 49, Saint Petersburg 197101, Russia}

\begin{abstract}
We show that the second harmonic generation (SHG) is enhanced in the chiral one-dimensional electron currents in a broad frequency range. The origin of the enhancement is  two-fold: first, the linear dispersion of the current and the associated plasmonic mode as well as the quasi-linear dispersion of plasmon-polariton result in the lift of the phase matching condition. Moreover, the strong field localization leads to the further increase of the SHG in the structure. The results suggest that the chiral currents localized at the domain walls of magnetic topological insulators can be an efficient source of second harmonic signal in the terahertz frequency range.
\end{abstract}

\maketitle
In recent years the field of plasmonics has been enjoying the exploration of plasmonic excitations in novel topological materials~\cite{lai2014plasmonics,stauber2014plasmonics,stauber2017plasmonics}. The appeal of the topological materials for  plasmonics is largely dictated by the fact, that the topologically protected surface currents in these materials support plasmonic excitations inheriting the immunity to backscattering, resulting in the suppression of the net plasmonic loss rate, which is of paramount importance for enabling applications of plasmonics in various fields~\cite{Boltasseva2011}. The field of topological plasmonics is now rapidly evolving and a plethora of novel low-loss plasmonic excitations has been predicted and observed in various topological insulators~\cite{Song4658,Kumar2016,Pietro2013,autore2015plasmon,Jin2016} and other topological materials, such as e.g. Weyl semimetals~\cite{Pellegrini2015,Hofman2016,song2017fermi,Polini2018}.

One of the most promising applications of plasmonics is the enhancement of the nonlinear optical processes, specifically second and higher harmonic generation~\cite{HHG1,HHG2,HHG3,de2016harmonics}. The amplification of the nonlinear signal is achieved due to the plasmon-assisted field enhancement. One of the limiting factor for the harmonic generation efficiency are the ubiquitous Ohmic losses. In this perspective, exploitation of topological plasmons with suppressed loss rates for the nonlinear frequency conversion could significantly enhance the conversion efficiencies.

Noteworthy, the topologically non-trivial photonic structures have been recently proposed for the enhancement of the higher harmonic generation (see the review ~\cite{smirnova2019nonlinear} and references within). While, in most of these studies the nonlinear current is produced by the conventional optically nonlinear media (such as lithium niobate or GaAs), the topologically non-trivial edge and surface states emerging in these structures facilitate the strong field enhancement, extended lifetime and unidirectional mode propagation which cumulatively increase the conversion efficiency.

At the same it has been shown in a number of papers that the linear electronic dispersion arising in topologically protected surface states as well as in low dimensional Dirac materials such as graphene may result in drastic enhancement of the nonlinear current~\cite{Mikhailov2011,glazov2011second,TopoSHG}.

Topological plasmon-polaritons, composite quasi-particle, a superposition of the topologically protected surface or edge current and an electromagnetic field could thus provide a two-fold source for the enhancement of the nonlinear signal, since they emerge due to the interaction of the electrons with linear dispersion, immune to backscattering and subwavelengthly localized electromagnetic field.

In this Letter, we exploit this simple idea by studying the second harmonic generation by the edge plasmon-polariton (EPP) localized at the domain wall in magnetic topological insulator (MTI).

\begin{figure}[H]
	\center{\includegraphics[width=1\linewidth]{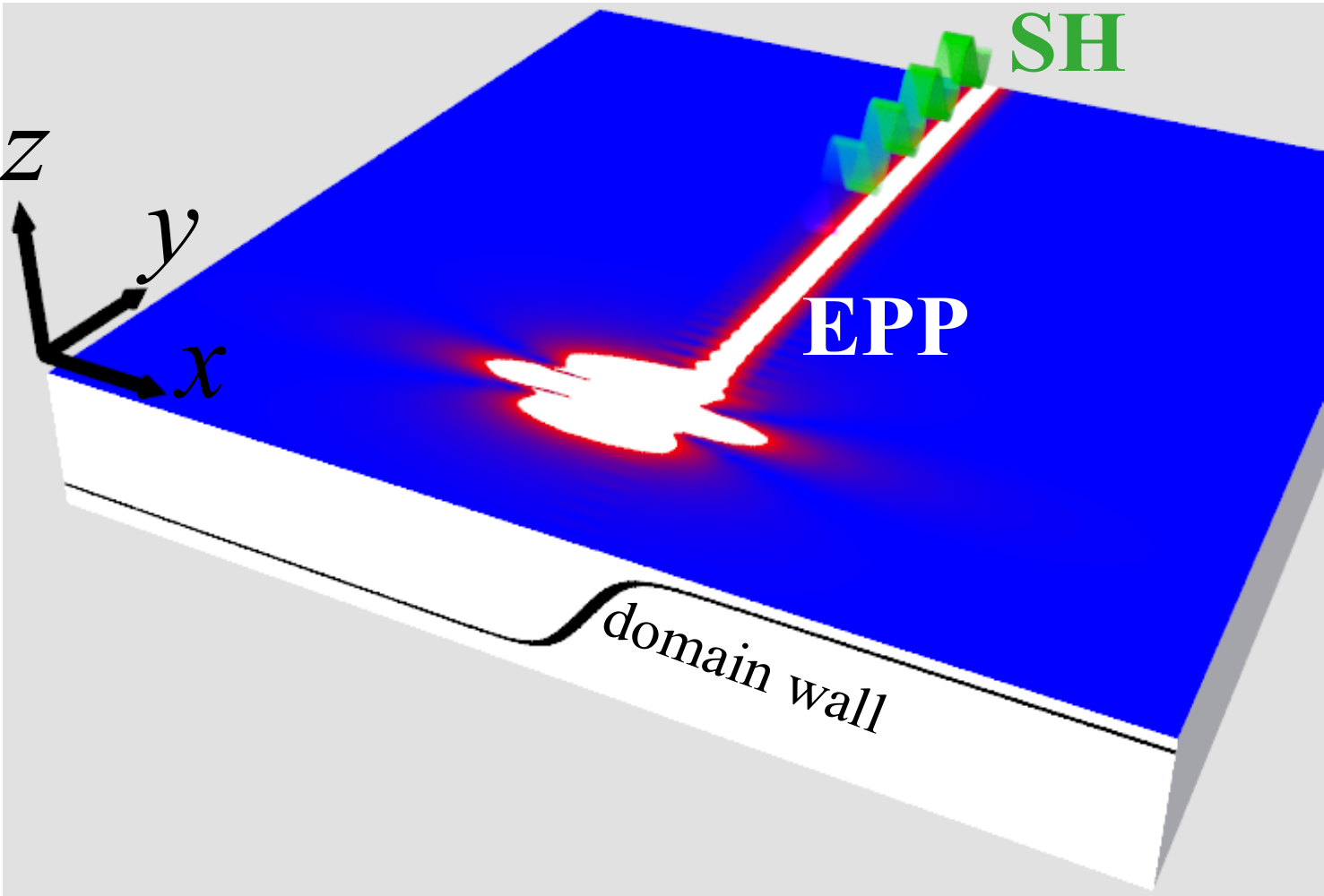}}
	\caption{Scheme of the SHG by the chiral current in magnetic TI. A point dipole excited an EPP localized at the domain wall. Nonlinear conductivity results in the emergence of the SH signal which is also localized at the domain wall}\label{fig01}
\end{figure}

MTI    can be realized, e.\,g., in the form of a ferromagnet thin film in a close proximity to a surface of 3D topological insulator or topological semimetal~\cite{EuS2014,Lee2016,Hu2017}. A perpendicular-to-the plane magnetization component in the ferromagnet induces a finite effective mass of the otherwise massless surface electrons. This results in a band gap in the spectrum of the surface states. 
A one-dimensional domain wall in the ferromagnet is, however, imaged in the Dirac electron system as a zero mass line that supports a helical electronic state. These quasi-one dimensional edge states are characterized by the linear electron dispersion and are Anomalous Hall counterparts of the Quantum Hall edge states.

In~\cite{ACS_helical} we have shown that these currents support a strongly localized low-loss helical plasmon polaritons with almost linear dispersion. Here, we consider a second harmonic generation suported by this EPP mode. Namely, we consider the situation shown schematically in Fig.~\ref{fig01}. A linear EPP is excited by a point-like scatterer (it may be a tip of scattering near-field optical microscop~\cite{Chen2012,Fei2012, Basov2016} or a deeply subwavelength resonant nanoantenna~\cite{NanoWirePlasmonics}). We then calculate the nonlinear conductivity, nonlinear current and the intensity of the second harmonic signal in this set-up.

The Hamiltonian of the single domain wall in the MTI structure is described by the Hamiltonian 
\begin{align}
    \hat{H}=v[\mathbf{\sigma}\times\mathbf{p}]_z+\Delta\tanh(x/a_0)\sigma_z,
\end{align}
where $v$ is the Fermi velocity,  $\Delta$ is the gap width proportional to the net magnetization, and $a_0$ is the width of the domain wall. The eigenergies of the edge states are given simply by $E=\hbar v k_y$, and the eigenstates are given by
\begin{align}
     \Psi_{\nu}(x,y)=F_{\nu}(x)\frac{e^{iqy}}{\sqrt{2\pi}},\quad F_{\nu}(x)=\frac{[a_0 \mathrm{B} (1/2,\nu)]^{-\frac{1}{2}}}{\cosh^{\nu}(x/a_0)}
\end{align}
where $\nu=a_0/l$, $l=\hbar v/\Delta$, and $\mathrm{B}$ is the Euler Beta function. In the limit of infinitely thin domain wall, $a_0~=~0$:
\begin{align}
    F(x)=F_{0}(x)=\frac{1}{\sqrt{l}}\exp\left[-\frac{|x|}{l}\right].
\end{align}

In what follows, we assume that the Fermi energy lies in the center of the bulk gap, and that the frequency of electromagnetic field is smaller then the gap width $\hbar\omega < \Delta$. Within this approximation we can neglect the excitation of the bulk states and assume that both linear and nonlinear current are generated solely by the intraband transitions of the edge state.
Both linear and nonlinear conductivities can be obtained within the unified formalism based on density matrix approach. Namely, the average current is given by
\begin{align}
    \langle \mathbf{J}(t)\rangle= \mathrm{Tr}[\mathbf{J}\rho(t)]=\sum_n \frac{e^{-\beta E_n}}{Z} \langle n(t) | \hat{\mathbf{J}}(t)| n(t) \rangle, \label{dens_mat}
\end{align}
where $\mathbf{J}$ is the current operator, $\rho(t)$ is the density matrix operator, $E_n$ and $|n(t)\rangle$ are the eigenvalues and eigenfunctions written in the interaction picture, respectively, and $Z$ is the corresponding partitition function.  The time-dependent eigenstates in the interaction picture are simply $|n(t)\rangle= e^{-i/\hbar \int ^t dt' V(t')} |n\rangle$, where the interaction term is given by
\begin{align}
    V(t) = -ev [\sigma \times \mathbf{A}(t)],
\end{align}
where $\mathbf{A}(t)$ is the vector potential of the perturbing field. The current operator is found as $\mathbf{J}=\partial{V}/\partial \mathbf{A}(t)$.
The linear conductivity of this system has been evaluated in~\cite{ACS_helical} and is written as:
\begin{align}
\sigma_{yy}(\omega,q,x,x',z,z')=\frac{\alpha i}{2\pi } \frac{v^2 q}{\omega} \frac{F_{\nu}^2(x)F_{\nu}^2(x')}{(k_0-v q/c)} \delta(z)\delta(z'), \label{eq_sigma_lin}
\end{align}
where $\alpha$ is the fine structure constant, $k_0=\omega/c$ and $F(x)$ is the function describing the transverse profile of the quasi-one dimensional current. In the limit of the infinitely narrow domain, $F(x)$ is given by

\begin{figure}[!h]
	\center{\includegraphics[width=1\linewidth]{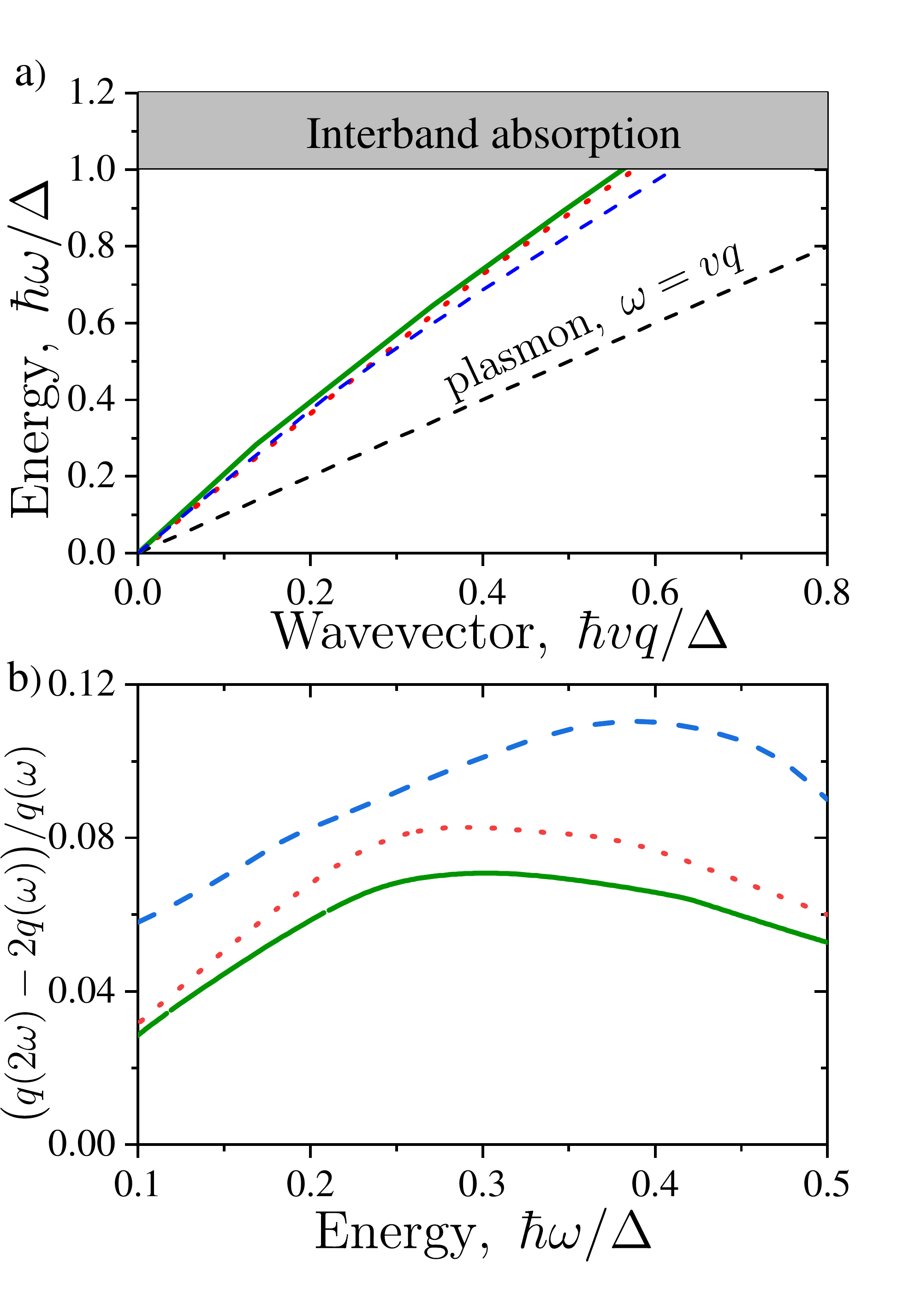}}
	\caption{(a) Dispersion of the EPP for different values of $\nu$: $\nu=0.1$ (solid green), $\nu=0.5$ (dotted red) and $\nu=1$ (dashed blue). (b) The dimensionless parameter of the phase mismatch. The line legend is the same for figure (a).}\label{fig2}
\end{figure}

It can be seen that the linear dispersion has a resonance at the dispersion of the chiral plasmon $\omega=vq$. While these plasmons can not be excited by a plane waves since their dispersion lies well below the light cone, they can be excited by a evanescent fields of the point-like scatterers. The dressing of the chiral plasmon by the electromagnetic field leads to formation of plasmon-polariton defined by the equation~\cite{ACS_helical}:
\begin{align}
&S(q,\omega)=\nonumber\\&{1-\tilde{v}\tilde{q}}-\alpha \tilde{v}^2\tilde{q}\frac{(\tilde{q}^2-1)}{\Gamma^4(\nu)}\int_0^{\infty}dx\frac{|\Gamma(\nu(1+ix/2))|^4}{\sqrt{q^2l^2+x^2-k_0^2l^2}}=0,  \label{lin_pol} 
\end{align}
which in the limit $a_0=0$ reduces to
\begin{align}
&S(q,\omega)=\nonumber\\&{1-\tilde{v}\tilde{q}}-\alpha \tilde{v}^2\tilde{q}(\tilde{q}^2-1)\frac{(1+\kappa^2)\tanh^{-1}\kappa-\kappa}{2\pi\kappa^3}=0,  \label{lin_pol} 
\end{align}
where $\tilde{v}=v/c$, $\tilde{q}=q/k_0$ and $\kappa^2=1+(\hbar\omega/2\Delta)^2\tilde{v}^2(1-\tilde{q}^2)$. 
The dispersion defined by Eq.~\eqref{lin_pol} is shown in Fig.~\ref{fig2}(a). 
We can see that the plasmon is weakly hybridised by the electromagnetic field and the dispersion of the plasmon-polariton is close to the one of plasmon. Moreover, we can see that for the reasonable values of $\nu$ the dispersion of EPP depends on $\nu$ only weakly. For the currently known experimental realizations of MTI the value of $\nu$ lies in the range $\nu\sim 10^{-2}-10^{-1}$. We can see that at $\nu=0.1$ the dispersion of EPP becomes indistinguishable from the one with $\nu=0$. In what follows we all assume $\nu=0$ in the calculations. Finally, it can be seen that within the gap the dispersion of the edge state is almost linear. Specifically, it becomes inreasingly linear as $\nu$ approaches zero.  Figure~\ref{fig2}(b) shows the quantity describing the linearity of the dispersion $(q(2\omega)-2q(\omega))/q(\omega)$. This quantity can be regarded as the dimensionless phase mismatch. It can be seen that as $\nu$ approaches zero the phase mismatch is small across all of the gap region
As  shown in \cite{ACS_helical} the structure excited by a point-like scaterrer such as a tip of scattering SNOM would support a long-living quasi one-dimensional plasmon polariton with the dispersion defined by~\eqref{lin_pol}. 

The nonlinear conductivity responsible for the second harmonic generation can be calculated straightforwardly from the expression~\eqref{dens_mat}. The nonlocal nonlinear conductivity is found from the relation
\begin{align}
  &  {J}_i(x,2\omega,2q)=\nonumber\\ \iint dx_1 dx_2 &\sigma_{ijk}^{SHG}(\omega,x,x_1,x_2,q)E_{\omega,j}(x_1,q)E_{\omega,k}(x_2,q) \label{JnlBasic}
\end{align}

The details of the calculation can be found in SI, and the expression for $\sigma_{nl}$ is given by:
\begin{align} 
    \sigma^{SHG}_{yyy}=\frac{c}{e}\tilde{q}\frac{\alpha^2\tilde{v}^3}{2\pi}\frac{ k_0^{-3}F^2(x)F^2(x_1)F^2(x_2)} {(1-\tilde{v}\widetilde{q})^2}
\end{align}
We see that according to the symmetry restrictions, since our system possesses the center of symmetry, the second harmonic current should be proportional to the wavevector of light in the direction of propagation~\cite{DurnevTarasenko}, $J(2\omega)\sim E^2\tilde{q}$.

In calculation of the linear and nonlinear conductivity we have neglected the processes of photoinization, i.e. the direct transitions between the edge states and the bulk states in the conduction and valence bands. This approximation is valid when (for the case of the Fermi energy in the center of the gap) $2\omega<\Delta$. Moreover, unlike the case of the spin Hall effect edge currents, there is only one edge state per edge, thus direct optical transitions between two edge states with opposite helicities do not occur. In order to avoid the thermal excitation of the bulk states we also consider the limit $T\ll \Delta$. The only source of the spreading of the electron wavepacket is thus the momentum relaxation of the chiral electrons in the channel due to impurity scattering which is weak in the quantum Hall edge currents.

We now consider the situation similar to one considered in~\cite{ACS_helical}: a helical edge plasmon polariton is excited by a point-like scatterer and propagates along the domain wall. At the sufficient distance from the scatterer the profile electric field is dominantly defined by the field of the edge plasmon-polariton (EPP). Its $y$ component, the only one responsible for the nonlinear current generation reads for the plane $z=0$
\begin{align}
    E_y^{EPP}(x,y)=E_0\mathcal{E}(\omega,q_{EP},x)e^{iq_{EP}y},
\end{align}
where $E_0$ is the amplitude defined by the coupling efficiency of the point-scatterer field to the EPP mode, $q_{EP}(\omega)$ is the wavevector of the EPP, defined by Eq.~\eqref{lin_pol}, and dimensionless functions $\mathcal{E}$ defines the profile of the field:
\begin{align}
    \mathcal{E}(\omega,q,x)=\int dx' G(x-x',\omega,q) F^2(x'),
\end{align}
where the Green's function $G$ is given by
\begin{align}
    G_{yy}(x-x',\omega,q)\sim\mathrm{K}_0(\sqrt{q^2-k_0^2}|x-x'|),
\end{align}
where $\mathrm{K}_0$ is the Macdonald function. 

\begin{figure}[!h]
	\center{\includegraphics[width=1\linewidth]{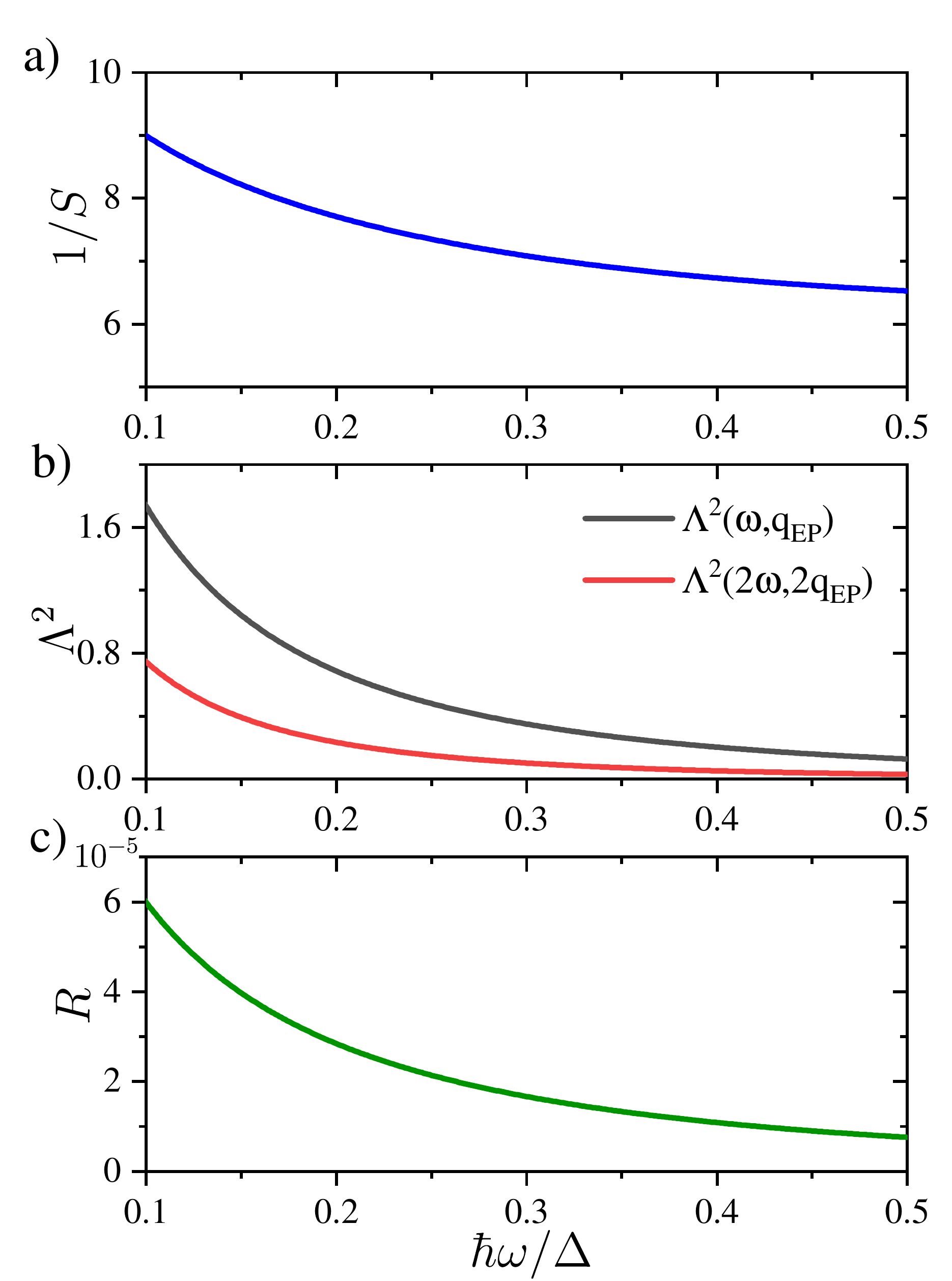}}
	\caption{Different contributions to the factor $R(\omega)$: (a)
	the effective phase-matching condition $1/S$; (b) Field localization at the fundamental and second harmonic frequencies ; (c) the spectrum of $R(\omega)$.}\label{fig3}
\end{figure}

According to Eq.~\eqref{JnlBasic} the nonlinear current can be written as
\begin{align}
    J_y(x,2\omega,y)=\frac{c}{e}\tilde{q}_{EP}\frac{\alpha^2\tilde{v}^3}{2\pi}\frac{ F^2(x)E_0^2\Lambda^2(\omega,q_{EP})} {k_0^3(1-\tilde{v}\tilde{q}_{EP})^2}e^{2iq_{EP}y},
\end{align}
where $\Lambda(\omega,q)=\int dx F^2(x) \mathcal{E}(\omega,q,x)$.
The bare field at second harmonic can then be written as
\begin{align}
E_{bare}(2\omega,x,y)=E_{2\omega}\mathcal{E}(2\omega,2q_{EP},x)
\end{align}
 However, the electric field at the second harmonic also gets renormalized due to the hybridization with the linear EPPs at the second harmonic. Collecting all the terms together, we get the expression for the electric field at the second harmonic
\begin{align}
    &E_y(x,2\omega,y)e^{-2iq_{EP}y}=R(\omega)\frac{\alpha \tilde{v}}{e k_0^2}E_0^2\mathcal{E}(2\omega,q_{EP},x)\label{EnlFull},
\end{align}
where
\begin{align}
    R(\omega)=\left[1+\frac{\alpha\tilde{v}\Lambda(2\omega,2q_{EP})}{S(2q_{EP},2\omega)}\right]\frac{2\tilde{q}_{EP}\alpha\tilde{v}^2\Lambda^2(\omega,q_{EP})} {(1-\tilde{v}\tilde{q}_{EP})^2} \label{eq:enhance}
\end{align}
where $S$ is defined by Eq.~\eqref{lin_pol}. Different terms entering Eq.~\ref{eq:enhance} are plotted in Fig.~\ref{fig3}. Namely, the term $S$ in the denominator  can be regarded as the phase-matching factor. Naturally, due to the almost linear dispersion of the EPP, $S$ is quite small, and the resonant contribution to the SHG signal is significant as shown in Fig.~ref{fig3}(a). The terms $\Lambda$ correspond to the field enhancement due to the subwavelength field localization in EPP mode. They are shown in Fig.~\ref{fig3}(b). The function $\Lambda$ has a logarithmic divergence in the limit of low frequencies. This however can be regularized either by introducing small but finite skin-depth of the edge current in $z$ direction or introducing a low frequency cut-off which is done further in the manuscript. We plot $R(\omega)$ in Fig.~\ref{fig3}(c). 

Omitting the spatial profiles, the ratio of the field amplitudes at the second and fundamental harmonic can be presented as
\begin{align}
    E_{SHG}/E_0=R(\omega) \left[\frac{eE_0v/\omega}{\hbar\omega}\right]. \label{eq:simple}
\end{align}
We can see that the efficiency of the SHG is proportional to the ratio of the maximum  kinetic energy gain by electron per the EM field period and the photon energy. First, let us recall that in the conventional conducting systems in the limit of low frequencies, the mean momentum is proportional to the relaxation time $\tau$ rather then to the field period. However, at low temperatures the anomalous Hall edge current can have extremely long relaxation times, and effect of its finiteness may be neglected. It is evident, that when the kinetic energy gain per EM cycle exceeds $\Delta$ the electrons reach the bulk conduction band and our approximations can not be applied. It means, that the upper limit for the numerator of the bracketed expression in Eq.~\eqref{eq:simple} is $\Delta$. At the same time, we have considered the infinitely long domain wall. This approximation holds, when the longitudinal extent of the domain wall $L$ is much larger than the effective wavelength of the plasmon polariton which can be approximated by $\lambda_{EP}\sim 2\pi v/\omega$. With this we can estimate the upper limit for the quantity in brackets.
\begin{align}
   Q(\omega)= \left[\frac{eE_0v/\omega}{\hbar\omega}\right]< \frac{\Delta L}{2\pi\hbar v}=\frac{L}{2\pi l}
\end{align}
For the domain wall length of $10$ microns, the dimensionless quantity can be as large 20.
For the adequate electric field amplitudes a more accurate approximation may be made. Namely, the characteristic time scale is defined not by momentum relaxation $\tau$ but rather by the time $\tau_0=L/v$ which takes electron to travel along  the whole domain wall. In this case we can write
\begin{align}
    Q(\omega)<\frac{eE_0L}{\Delta}\frac{L}{2\pi l}
\end{align}
Let us consider the specific case of $\hbar\omega=0.5$ meV, $\Delta=1.5$ meV, and $v=c/600$. In this case, $l=\approx 160$ nm. We also assume $L=10$ microns. In this case, $\tau_0\approx 20 ps$ and $T=2\pi/\omega\approx 10 \mathrm{ps} \approx \tau_0$. In this case, for moderate field amplitude of $1\mathrm{V}/\mathrm{cm}$, $Q(\omega)\approx 6.5$. Multiplying it by the respective value of $R$ we can estimate the effective nonlinear susceptibility as $10^{-4}~\mathrm{cm V^{-1}}$ which is almost four orders of magnitude larger than in the bulk $GaAs$, $1.7\times 10^{-8}~\mathrm{cm V^{-1}} $~\cite{nelson2000nonlinear}.  This suggest that  the chiral currents may be  regarded as an  extremely efficient source of second harmonic generation in terahertz range. 

We stress that the aforementioned response is broadband and does not require any additional photonic resonant structure, while it is evident that the latter would further increase the SHG signal. In the estimation of the effective nonlinear response we did not account for the efficiency of coupling of the fundamental harmonic signal to the EPP mode, which is usually week due to the strong localization of the EPP. At the same, the are now many routes of efficient coupling of the bulk field to the deeply subwavelength plasmonic terahertz modes~\cite{bahk2019terahertz}. It is also noteworthy that the broad band response is achieved due to the almost linear dispersion of EPPs in the structure providing the lift of the strict phase-matching conditions.

To conclude, we have considered the second harmonic generation in the chiral current localized at the domain wall of magnetic topological insulator. Assisted by the excitation of the edge-plasmon polariton both at fundamental and second harmonic frequency, the SHG process can be several orders more efficient than in bulk $GaAs$. The effect is broad-band due to the linear dispersion of both the current and plasmon-polariton mode, and due to the absence of the backscattering in the chiral current, its magnitude is virtually limited only by the domain wall length. Thus, we anticipate, that the nanostructures comprising domain walls in MTI can become a building block for the efficient sources of SHG in terahertz range.

\bibliography{TopMagn}
\end{document}